\documentclass[12pt,a4paper]{article}
\usepackage{jheppub}
\usepackage{amsmath,amssymb,epsf}
\usepackage{graphicx,color}
\usepackage{amssymb}
\usepackage{amsmath}
\usepackage{mathrsfs}
\usepackage{graphics}
\usepackage{epsfig}
\usepackage{graphicx}
\usepackage{subfigure}
\usepackage{bm}
\usepackage{nicefrac}
\usepackage{feynmp}

\def\beq{\begin{equation}}
\def\eeq{\end{equation}}
\def\baq{\begin{eqnarray}}
\def\eaq{\end{eqnarray}}

\def\be{\begin{equation}}
\def\ee{\end{equation}}
\def\bea{\begin{eqnarray}}
\def\eea{\end{eqnarray}}

\def\mp{M_{P}}

\def\k{{\bf k}}

\def\g{\gamma}

\def\k{\kappa}                    
\def\l{\lambda}

\def\6{\partial}

\newcommand{\bright}{\begin{flushright}}
\newcommand{\eright}{\end{flushright}}
\newcommand{\bminip}{\begin{minipage}}
\newcommand{\eminip}{\end{minipage}}
\newcommand{\bcent}{\begin{center}}
\newcommand{\ecent}{\end{center}}

\title{Non-thermal Gravitino Production in Tribrid Inflation}
\author[a,b]{Stefan Antusch}
\author[c]{Koushik Dutta}
\affiliation[a] {Department of Physics, University of Basel,
Klingelbergstr. 82, CH-4056 \\Basel, Switzerland}
\affiliation[b]{Max-Planck-Institut f\"ur Physik (Werner-Heisenberg-Institut), F\"ohringer Ring 6,\\ D-80805 M\"unchen, Germany}
\affiliation[c] {Theory Division, Saha Institute of Nuclear Physics, 1/AF Saltlake, Kolkata, India}
\emailAdd{stefan.antusch@unibas.ch}
\emailAdd{koushik.dutta@saha.ac.in}
\abstract{
We investigate non-thermal gravitino production after tribrid inflation in supergravity, which is a variant of supersymmetric hybrid inflation where three fields are involved in the inflationary model and where the inflaton field resides in the matter sector of the theory. In contrast to conventional supersymmetric hybrid inflation, where non-thermal gravitino production imposes severe constraints on the inflationary model, we find that the ``non-thermal gravitino problem'' is generically absent in models of tribrid inflation,
mainly due to two effects: (i) With the inflaton in tribrid inflation (after inflation) being lighter than the waterfall field, the latter has a second decay channel with a much larger rate than for the decay into gravitinos. This reduces the branching ratio for the decay of the waterfall field into gravitinos. (ii) The inflaton generically decays later than the waterfall field, and does not produce gravitinos when it decays. This leads to a dilution of the gravitino population from the decays of the waterfall field. The combination of both effects generically leads to a strongly reduced gravitino production in tribrid inflation.

}

\begin{document}
\maketitle

\section{Introduction}
 
Cosmic inflation is a successful paradigm for the early universe. Among the various types of inflationary theories proposed, ``hybrid-like'' models \cite{Linde:1991km} offer the attractive possibility to connect them to a particle physics phase transition around energy scales $M_\mathrm{GUT}$ of Grand Unification. In such models, inflation proceeds along a nearly flat valley in field space with large vacuum energy. When a ``critical value'' is passed, inflation is ended by a second order phase transition where the so-called waterfall field moves quickly towards its minimum. 

In the ``conventional'' realisation of hybrid inflation in supersymmetry (SUSY) or supergravity (SUGRA) \cite{Dvali:1994ms,Copeland:1994vg,Linde:1997sj}, the waterfall field can be associated with the spontaneous breaking of some symmetry around $M_\mathrm{GUT}$, however the inflaton field  is necessarily a singlet under any  symmetry (apart from possible R-symmetries). It is therefore somewhat decoupled from the rest of the particle theory. 

This can be overcome in tribrid inflation \cite{Antusch:2008pn,Antusch:2009vg}. In tribrid inflation, three fields are involved in the inflationary model: the inflaton field, the waterfall field, and a so-called ``driving field'' (which however does not contribute to the dynamics during and after inflation). In contrast to ``conventional'' hybrid inflation, the inflaton field can now reside in the matter sector of the theory. One can, for example, identify it with one of the right-handed sneutrinos \cite{Antusch:2004hd} or with a D-flat direction of fields in representations of the gauge symmetry \cite{Antusch:2010va}. This allows for particularly close connections between inflation and the underlying particle physics theory (see e.g.\ \cite{Antusch:2010mv,Antusch:2013toa}).

Inflationary models of the early universe face various constraints: For instance, they have to be consistent with the data on primordial fluctuations extracted from CMB observations, which requires a comparatively flat potential during inflation. After inflation, the models have to successfully reheat the universe and produce a thermal bath of Standard Model particles. In addition, not too many long lived cosmic relics should be produced. One example of the latter problem in the context of supergravity is the so-called ``gravitino problem''  \cite{Weinberg:1982zq}.

Gravitinos are the superpartners of the gravitons in SUGRA. Since they interact quite weakly (with strength of gravitational interactions) they are typically rather long lived and can be problematic in various respects: In particular, if they are stable, they can easily overclose the universe. When they are unstable and decay after BBN (or around the time of BBN), their energetic decay products can destroy the produced light elements. But also if they decay before BBN, which is typically the case when they are heavier than about 30 TeV, they impose significant constraints. The decay of each gravitino (assuming R-parity conservation) in general produces one stable lightest supersymmetric particle (LSP), which would count towards the dark matter (DM) relic abundance. The constraint on the number density of gravitinos in this case comes from the requirement not to overproduce DM. 

Gravitinos can be produced thermally, in the thermal bath after reheating \cite{Bolz:2000fu,Pradler:2006qh,Pradler:2006hh}, and non-thermally from the decays of the fields involved in the dynamics after inflation \cite{Kawasaki:2006gs,Kawasaki:2006hm}. The thermal production imposes constraints on the reheat temperature. For instance, assuming a gravitino with mass above 30 TeV (with conserved R-parity) and a LSP mass of about 100 GeV, this bound is roughly $T_{R} \lesssim \mathcal{O}(10^{10}\:\,\text{GeV})$ (see, e.g., \cite{Kawasaki:2004qu}). For much smaller $T_{R}$, the thermal production can be neglected in this example case. On the other hand, it is known that for the case of ``conventional'' SUSY hybrid inflation the non-thermal production of gravitinos provides a potential  problem \cite{Kawasaki:2006hm}. 

In this work, we therefore investigate non-thermal gravitino production after tribrid inflation and compare the results to the case of ``conventional'' SUSY hybrid inflation. We find that the ``non-thermal gravitino problem'' is generically absent in tribrid inflation, mainly due to two effects: 
(i) With the inflaton being lighter than the waterfall field, the latter automatically has a second decay channel with a much larger rate than for the decay into gravitinos. 
(ii) The inflaton generically decays later than the waterfall field, and does not produce gravitinos when it decays. This leads to a dilution of the gravitino population from waterfall field decays. 
The combination of both these effects generically leads to a strongly reduced gravitino production in tribrid inflation.

In the next section, Sec.~\ref{Tribrid_inflation_vs_conventional_SUSY_hybrid_inflation}, we will review ``conventional'' SUSY hybrid inflation and tribrid inflation, with a focus on the post-inflationary dynamics. In Sec.~\ref{Non-thermal_gravitino_production_in_SUGRA}, we discuss the non-thermal gravitino production from a generic scalar field at the end of inflation, followed by the calculation of the gravitino yield in ``conventional'' SUSY hybrid inflation in Sec.~\ref{Non-thermal_gravitino_production_in_hybrid_inflation}. In Sec.~\ref{grav_production_tribrid}, we calculate the gravitino production in tribrid inflation and discuss our results. Our conclusions are presented in Sec.~\ref{conclusion}.  

\section{Tribrid inflation and conventional SUSY hybrid inflation} \label{Tribrid_inflation_vs_conventional_SUSY_hybrid_inflation}

There are two classes of supersymmetric hybrid-like F-term inflation models: ``Conventional'' F-term SUSY hybrid inflation \cite{Dvali:1994ms}  and tribrid inflation \cite{Antusch:2009vg}. Both have in common that there is an inflaton field responsible for the dynamics during inflation and a waterfall field responsible for ending inflation. The end of inflation may be associated with a particle physics phase transition, which opens up the possibility to embed inflation in particle physics models such that the combined theory becomes more predictive and testable. 

\vspace{2mm}
	{\bf``Conventional'' F-term hybrid inflation:} When SUSY (or SUGRA) hybrid inflation is discussed, the model is usually described by the superpotential
\be
	W = \k \, \phi \, (H^2 - M^{2}) \, ,
\ee
combined with a (at least nearly) canonical K\"ahler potential \cite{Dvali:1994ms,Copeland:1994vg,Linde:1997sj}. The superfield $\phi$ contains the inflaton field as a scalar component. For simplicity, we have assumed $H$ to be a singlet, but $H^2$ can be replaced by $H\bar H$ where $H$ and $\bar H$ belong to the conjugate representation of a gauge group. We will use the same symbols for the superfield and the scalar component throughout the paper.

In conventional F-term hybrid inflation, $\phi$ is a singlet\footnote{It may only be charged under some R symmetry.}. $\phi$ is the so-called ``driving field'' which generates the potential for $H$ by its F-term $| F_{\phi}|^2$, but it is usually somewhat disconnected from the rest of the particle physics theory. During inflation, $\langle H \rangle = 0$  and $\phi \neq 0$ (false vacuum), driving inflation. The tree-level inflationary potential $V_0 = \k^2 M^4$ is corrected by quantum corrections (and K\"ahler potential corrections) generating the desired slope of the potential. Without loss of generality we consider real $\k$ and $M$. With $\kappa \sim 0.1$, an appropriate matching of the amplitude $A_s$ of the scalar fluctuations requires $M \sim 5\cdot 10^{15}$ GeV. On the other hand, an exactly canonical K\"ahler potential would be inconsistent with the PLANCK observation of the spectral index \cite{Planck:2015xua}. This tension can be resolved, for instance, with higher order terms in the 
K\"ahler potential for the inflation field \cite{BasteroGil:2006cm}, or by considering the effects of soft supersymmetry breaking terms \cite{Rehman:2009nq}.

$H$ is the waterfall field which ends inflation by a second order phase transition when $\phi$ rolls below a critical value where the mass of $H$ gets tachyonic. After this ``waterfall'' (assuming homogenous fields) both fields, the inflaton and the waterfall field, oscillate around their minima. When the SUSY breaking effects are neglected, the minima are at $\langle \phi \rangle = 0$, and $\langle H \rangle = M$.
Including SUSY breaking effects after inflation, the waterfall field and the inflaton field mix almost maximally and are almost degenerate in mass, 
\begin{equation} \label{hybrid_mass}
\phi_{\pm} = \frac{1}{\sqrt{2}}(\phi \pm H)~,
\end{equation}
with $m^2_{\phi_\pm} \simeq 4 \kappa^2 M^2$. Let us now turn to tribrid inflation, which in several respects has quite different features.
	
\vspace{2mm}
	{\bf ``Tribrid'' inflation:}  In tribrid inflation \cite{Antusch:2008pn,Antusch:2009vg}, the inflaton can be associated with a matter field of the underlying particle physics theory. Here, for simplicity, we consider the case that it is a singlet as well (e.g.\ one of the right-handed sneutrinos in extensions of the MSSM to explain the observed neutrino masses) but it may also be a D-flat direction of gauge non-singlet matter fields as discussed in \cite{Antusch:2010va,Antusch:2011ei}. There are several variants of tribrid inflation in supergravity \cite{Antusch:2008pn,Antusch:2012jc,Antusch:2012bp}, but we will focus here on a simple representative where (in Planck units) the superpotential has the form\footnote{A more general form reads $W = \k \, X \, ( H^{\ell}  - M^{2} ) + \l \,  H^{m}\,\phi^{n}$, with $\ell,m,n \ge 2$ \cite{Antusch:2012jc,Antusch:2012bp}.} 
\be \label{eq:W_tribrid}
	W = \k \, X \, ( H^2  - M^{2} ) + \l \,  H^2\,\phi^{2} \, .
\ee	
The driving field is now called $X$. It is not the inflaton and only provides the potential for the waterfall field $H$ by ist F-term $|F_{X}|^2$. By a term $\gamma_X |X|^4$ in the K\"ahler potential $K$ with $\gamma_X < -1/3$ it will have a mass larger than the Hubble scale during inflation and thus stays at $\langle X \rangle = 0$, not affecting the inflationary dynamics. We note that during inflation $W = W_{\phi} = 0$ for tribrid inflation, which e.g.\ has the nice feature that it allows to combine low scale supersymmetry breaking with high scale inflation, as discussed in \cite{Mooij:2010cs,Antusch:2011wu}.

The superfield $\phi$ contains the inflaton as scalar component. The second term in $W$ generates a supersymmetric mass term for $\phi$ at the end of inflation when $\langle H \rangle \neq 0$. During inflation, the inflaton potential would be flat at tree level and in global SUSY (with the vacuum energy during inflation given by $V_0 = \k^2 M^4$). The desired slope of the potential is generated by loop corrections (with SUSY broken during inflation due to the non-zero $F_{X}$) or from higher-dimensional effective operators in $K$. Consistency of the amplitude $A_s$ of the scalar perturbations and of the value of the spectral index $n_s$ with observations can be achieved, e.g., with the parameter choices $ M \sim 5\cdot 10^{15}$ GeV, $\kappa \sim 0.1$, and $\lambda \sim 0.1$ \cite{Antusch:2009ef}. These values are chosen to match the set of viable parameter values given above for conventional SUSY hybrid inflation models. When we estimate the gravitino production in both classes of inflation models, we will use these values as benchmark point.

After inflation, we have to consider the dynamics of three fields $X,H$ and $\phi$. $X$ and $H$ mix almost maximally (when the SUSY breking effects after inflation are included) and the relevant mass eigenstates are
\begin{equation}
\chi_{\pm} = \frac{1}{\sqrt{2}} (X \pm H)\:,
\end{equation} 
with nearly degenerate masses $m_{\chi_{\pm}}^2 \simeq 4 \kappa^2 M^2$. The minima are at $\langle X \rangle = 0$, and $\langle H \rangle = M$.

The inflaton field $\phi$ has its potential minimum at $\langle \phi \rangle = 0$, and this is true even after including SUSY breaking effects. At the minimum the inflaton mass is given by $m_{\phi}^2 \sim 4\lambda^2 M^4/M_{P}^2$, with $\mp$ being the reduced Planck mass $\mp = 2.4 \cdot 10^{18}$ GeV. As we see, it is suppressed with respect to the mass $m_{\chi_{\pm}}^2$ of the mixed $X, H$ fields by $M^2/M_{P}^2$. Furthermore, due to the second term in $W$ of Eq.~(\ref{eq:W_tribrid}), the $\chi_{\pm}$ fields can decay into a pair of $\phi$ components, as we will discuss in more detail below.  

In summary, we can say that the dynamics at the end of inflation for the case of ``conventional'' SUSY hybrid inflation is governed by the oscillations and the decay products of the $\varphi_{\pm}$ field. For the case of tribrid inflation, both $\chi_{\pm}$ and $\phi$ will oscillate, and $\chi_{\pm}$ will decay first (dominantly into $\phi$ components plus some gravitinos), followed by the decay of the inflaton field $\phi$. The goal of this paper is to calculate the gravitino production from the decays of these fields.

\vspace{2mm}
	{\bf SUSY breaking sector:}
As we are going to discuss in the next section, the decay of the inflaton into gravitinos is caused by the mixing between the inflaton and the superfield $Z$ which after inflation breaks SUSY via its F-term. We will assume for simplicity that there is only one such field, that R parity is conserved, and that the gravitinos have a comparatively large mass, $> 30$ TeV, such that they decay before BBN. 

Under these conditions, the gravitino decays will not spoil BBN. On the other hand, each gravitino will in general produce one LSP which contributes to the dark matter abundance. This leads to a constraint from not overproducing dark matter, or one may even explain the dark matter abundance from this non-thermal production. Since we are considering small field inflation models in this work, we will not specify a particular form of the K\"ahler potential, but rather allow for a general expansion in terms of fields/$M_{Pl}$. We will include effective operators in the K\"ahler potential, such as $\gamma_X |X|^4$ mentioned above, or the ones which will be mentioned below.  

Let us discuss the role of the $Z$ field after inflation: Its fermionic component gets eaten up by the gravitino in the super-Higgs mechanism to give the gravitino its mass. The scalar component $Z$, on the other hand, remains in the theory and in general also adds to the dynamics. When it is displaced from its true minimum at the end of inflation and if it has mass and decay rate of sizes similar to the ones of the gravitino, it can also lead to a severe ``modulus problem'' \cite{Endo:2006zj,Nakamura:2006uc}. It may dominate the universe at late times and when it decays too late it can again spoil the predictions of BBN or, even when it decays before BBN, its decay into two gravitinos would reintroduce the gravitino problem. 

However, the problem can be overcome easily with terms  $- \g_{ZZ} \lvert Z \rvert^{4}/\mp^{2}$ and $-\g_{ZX} \lvert X \rvert^{2} \lvert Z \rvert^{2}/\mp^{2} $ present in the K\"ahler potential. $\g_{ZZ}$ increases the mass of the $Z$-modulus field, such that $m_Z \gg m_{3/2}$. Furthermore, with $\g_{ZX} \gtrsim 10$  the $Z$-modulus adiabatically tracks the minimum of the potential, such that only negligible amount of energy is transferred into $Z$ and  the modulus problem is solved \cite{Linde:1996cx,Nakayama:2011wqa}. In the following we will assume suitable K\"ahler potential terms $\g_{X}$, $\g_{ZZ}$ and $\g_{ZX}$, such that we can ignore the dynamics of $Z$ and $X$ and that after inflation $m_{\phi} \gg m_{Z} \gg m_{3/2}$ is satisfied.

\section{Non-thermal gravitino production in SUGRA} \label{Non-thermal_gravitino_production_in_SUGRA}

As mentioned above, the gravitino can be produced either thermally or non-thermally. The former production takes place in the thermal bath after reheating at the end of inflation, and its abundance increases with the reheat temperature $T_R$. In addition, gravitinos can also be produced non-thermally, e.g.\ from the decay of the inflaton field. 

The gravitino abundance might pose a threat to inflation models: For instance, when the gravitinos are unstable and decay too late, they may spoil the predictions of BBN. However even when they decay before BBN, which is typically the case when they are heavier than about 30 TeV, they impose a serious constraint. Assuming R parity is conserved, each gravitino decay produces one LSP, which adds to the DM abundance, and it must not exceed $\Omega_{DM}h^2 = 0.12$ \cite{Planck:2015xua}. This results in the constraint
\begin{equation}
m_{LSP} Y_{3/2}^{\varphi} = m_{LSP} \left(\frac{n_{LSP}}{s}\right) = \frac{\rho_{{\small LSP}}}{s} \lesssim \frac{\rho_{DM}}{s} = \Omega_{DM} \frac{\rho_c}{s} = 4.2\cdot 10^{-10}~\text{GeV}\:,
\label{gravitino_constraint}
\end{equation}
where $Y_{3/2}^{\varphi}$ is the gravitino to entropy ratio produced non-thermally from the decay of some scalar field $\varphi$ and $m_{LSP}$ is the mass of LSP. We have used the present critical density $\rho_c = 1.88\cdot 10^{-29}~h^2~g~cm^{-3}$, and the entropy density $s = 2969~cm^{-3}$. For our discussion, we neglect the thermally produced gravitino abundance (i.e.\ $Y_{3/2}^{thermal}$), which is justified for sufficiently small $T_R$ (i.e.\ typically $T_R \ll 10^{10}$ GeV).
We will consider the case that DM mainly consists of thermally produced LSPs. To derive a bound on the gravitino abundance, we nevertheless require that the non-thermally produced dark matter density does not exceed $\Omega_{DM}h^2 = 0.12$. Furthermore we will neglect annihilations of the non-thermally produced LSPs, which is justified for an LSP which can yield the correct thermal relic density.

Let us now turn to the non-thermal production: The part of the Lagrangian density which governs the decay of any scalar field $\varphi$ into a pair of gravitinos is (see e.g.\ \cite{Wess:1992cp})
\begin{equation}
e^{-1}\! \mathcal{L} = - \frac{1}{8}\epsilon^{\mu\nu\rho\sigma}(G_{\varphi}\partial_{\rho} \varphi + G_{Z}\partial_{\rho} Z - {\text H.c})\bar \psi_{\mu}\gamma_{\nu}\psi_{\sigma} - \frac{1}{8} e^{G/2}(G_{\varphi} \varphi + G_{Z} Z + {\text H.c})\bar \psi_{\mu}[\gamma^{\mu}, \gamma^{\nu}]\psi_{\nu}\,.
\end{equation}
Here $Z$ is the field which breaks SUSY by its F-term (and we assume only one field here for simplicity), $\psi_{\mu}$ is the gravitino field and $G \equiv K + \ln |W|^2$ with $K$ and $W$ being the K\"ahler potential and superpotential, respectively. In $G_\varphi$ and $G_Z$, the subscript denotes the derivative with respect to the field $\varphi$ or $Z$. 

The field $Z$ can mix with $\varphi$, and this mixing induces the decay of $\varphi$ into gravitinos, as discussed e.g.\ in  \cite{Dine:2006ii,Endo:2006tf}. The real and the imaginary components of the scalar field have the same decay rates to two gravitinos, given by
\begin{equation}\label{Gamma32_G_effective}
\Gamma_{3/2}^{\varphi}\equiv  \Gamma(\varphi \rightarrow 2\psi_{3/2}) \simeq \frac{|G_{\Phi}^{\text{eff}}|^2}{288\pi} \frac{m_{\varphi}^5}{m_{3/2}^2 M_{Pl}^2},
\end{equation}
where $|G_{\Phi}^{\text{eff}}|^2 := |G_{\varphi}|^2 + |\tilde \epsilon^* G_{\bar Z}|^2$ (following the notation of e.g.\ \cite{Endo:2006tf}), and $\tilde \epsilon^*$ is related to the mixing angle between the relevant scalar field and the SUSY breaking field. We have assumed that the inflaton mass (after inflation) satisfies $m_{\varphi} \gg m_{3/2}$. In the above expression, the only model dependent parameter is $G_{\Phi}^{\text{eff}}$ and for the considered case, $m_{\varphi} \gg m_Z \gg m_{3/2}$, it is given (in Planck units) by \cite{Endo:2006tf}
\begin{equation}
|G_{\Phi}^{\text{eff}}|^2 \simeq  \left|\sqrt{3} g_{\varphi\bar Z} \frac{m_Z^2}{m_{\varphi}^2}\right|^2 + \left|  \sqrt{3} (\nabla_{\varphi} G_Z) \frac{m_{3/2} m_Z^2}{m_{\varphi}^3}\right|^2 + \left|3 g_{\bar \varphi ZZ}\frac{m_{3/2}}{m_{\varphi}}\right|^2.
\label{G_effective}
\end{equation}
Here $g_{i\bar j}$ denotes derivatives of the K\"ahler potential with respect to the field $\varphi_i$ and $\bar \varphi_j$.
The first term in Eq.~(\ref{G_effective}) thus contributes whenever there is non-zero mixing between $\varphi$ and $Z$ in the K\"ahler metric.  
When $\varphi$ is charged differently then $Z$ (or $\overline Z$) under some symmetry (including R symmetry), then this term is always proportional to $\langle \varphi \rangle$.
Regarding the second term, we note that $|\nabla_{\varphi} G_Z| \sim \langle \varphi \rangle$ when the K\"ahler potential contains a canonical term for $\varphi$, $K = |\varphi|^2 + \dots $ \cite{Endo:2006tf}. The third term proportional to $g_{\bar \varphi ZZ}$ can contribute significantly to the decay rate as long as the SUSY breaking field is uncharged such that terms like $ a | \varphi |^2 (Z^2 + \overline{Z}^2)/M_{P}^2$ exist in $K$. The contribution to the decay rate from this term is then also proportional to the field vev $\langle \varphi \rangle$. 

In the following, we will apply these results to the fields which are relevant after inflation, which in ``conventional'' SUSY hybrid and tribrid inflation models are (primarily) the inflaton field and the waterfall field. Such fields can acquire non-zero values of $G_{\Phi}^{\text{eff}}$ from the terms in Eq.~(\ref{G_effective}), and then produce gravitinos when they decay. In general all three of these terms may contribute, however in order to keep the discussion simple we will focus on the effect of the last term of Eq.~\eqref{G_effective} and compare the resulting gravitino production in hybrid and tribrid inflation models. We choose this case as an example since when the third term is non-zero, it typically provides the dominant contribution to $G_{\Phi}^{\text{eff}}$.

\section{Non-thermal gravitino production in hybrid inflation} \label{Non-thermal_gravitino_production_in_hybrid_inflation}
As mentioned above, we will focus on the case that the contribution from the third term in Eq.~\eqref{G_effective} dominates the decay rates to gravitinos. More explicitly, we consider as example the effects of a K\"ahler potential term $\delta K = a |H|^2 (Z^2 + \overline{Z}^2)/M_{P}^2$, which exists when $Z$ is uncharged. We then obtain 
\begin{equation}
|G_{\phi_{\pm}}^{\text{eff}}| \simeq 3 a \frac{\langle H \rangle}{M_{P}} \frac{m_{3/2}}{m_{\phi_{\pm}}},
\end{equation}
and the decay rates of the mass eigenstates to gravitinos are given by\footnote{We note that $\Gamma_{3/2}^{\phi_{+}} =  \Gamma_{3/2}^{\phi_{-}} =: \Gamma_{3/2}^{\phi}$. We will nevertheless continue to use $m_{\phi_{\pm}}$ to distinguish it from the post-inflationary inflaton mass $m_{\phi}$ in tribrid inflation. 
}
\begin{equation} \label{gamma}
\Gamma_{3/2}^{\phi} = \frac{9 a^2}{288 \pi}\frac{\langle H\rangle^2 m_{\phi_{\pm}}^3}{M_{P}^4}.
\end{equation}

The $\phi_{\pm}$ fields decay mainly to the SM particles and their superpartners (with a total decay rate $\Gamma^{\phi}_{tot}$), giving rise to the reheat temperature 
\begin{equation}\label{eq:TR_Phi}
T_{R}^{\phi} = \sqrt{\Gamma^{\phi}_{tot} M_{P}} \left(\frac{\pi^2 g_{*}}{10}\right)^{-1/4},
\end{equation}
where $g_{*}$ is the effective number of relativistic degrees of freedom at the temperature $T_R$. Counting supersymmetric degrees of freedom, we take $g_{*}\sim228$. 
Neglecting gravitino production in the thermal bath, the (non-thermally produced) gravitino number density from $\phi$ decays $Y^\phi_{3/2}$, normalized to entropy density, is given by 
\begin{equation}\label{yield_basic}
Y_{3/2}^{\phi} \simeq 2 \frac{\Gamma_{3/2}^{\phi}}{\Gamma^{\phi}_{tot}}\frac{3}{4} \frac{T_{R}^{\phi}}{m_{\phi_{\pm}}} = \frac{3}{2} \left(\frac{\pi^2 g_{*}}{10}\right)^{-\frac{1}{2}} \frac{\Gamma_{3/2}^{\phi}}{T_{R}^{\phi}} \frac{M_{P}}{m_{\phi_{\pm}}}.
\end{equation}
Using Eq.~\eqref{gamma}, this leads to 
\begin{equation} \label{yield_hybrid}
m_{LSP} Y_{3/2}^{\phi} \simeq \kappa^2 a^2 \left(\frac{\langle H \rangle}{10^{16}~\text{GeV}} \right)^2\! \left(\frac{M}{5\cdot10^{15}~\text{GeV}} \right)^2\! \left(\frac{m_{LSP}}{100~\text{GeV}}\right) \!\left(\frac{10^{7}~\text{GeV}}{T_{R}^{\phi}} \right)10~\text{GeV},
\end{equation}
where we have used $m^2_{\phi_\pm} \simeq 4 \kappa^2 M^2$. Assuming $a \sim \mathcal{O} (1)$, and for typical values of $\langle H \rangle \sim 10^{16}~ \text{GeV}$, $M \sim 5\cdot 10^{15}~\text{GeV}$, $\kappa \sim 0.1$ \cite{BasteroGil:2006cm}, we see immediately that the estimate violates the observational constraint of Eq.~$\eqref{gravitino_constraint}$ for a wide range of parameters. Unless the higher order terms in the K\"ahler potential are highly suppressed, say $a \sim 10^ {-5}$, the gravitino production is problematic for ``conventional'' SUSY hybrid inflation models \footnote {One way to avoid the gravitino problem in ``conventional'' SUSY hybrid inflation arising from the third term of Eq.~\eqref{G_effective} is by charging $Z$ under a symmetry such that $g_{\bar \Phi ZZ}$ vanishes. Such a case has been studied, e.g., in \cite{Nakayama:2012hy}.}. While the thermal production of gravitinos can be reduced by lowering $T^\phi_R$, this would increase the non-thermal gravitino production in ``conventional'' SUSY hybrid inflation.

\section{Non-thermal gravitino production in tribrid inflation} \label{grav_production_tribrid}

In this section we estimate how many gravitinos are produced non-thermally in tribrid inflation, i.e.\ from the decay products of the inflaton $\phi$, the waterfall field $H$, and the ``driving field'' $X$. As it has been mentioned earlier, at the end of inflation $H$ and $X$ mix maximally to form the mass eigenstates $\chi_{\pm}$ (with nearly degenerate masses). Therefore, the dynamics of $\chi_{\pm}$ and $\phi$ is relevant for calculating the non-thermally produced gravitino abundance. 

\subsection{Decay rates in gravitinos} \label{Decay_rates_into_gravitinos}
As it has been noted in Eq.~\eqref{Gamma32_G_effective}, the decay rate to gravitinos from a scalar field depends on $G_{\Phi}^{\text{eff}}$ (given by Eq.~\eqref{G_effective}. For tribrid inflation $G_{\phi}^{\text{eff}}$ vanishes since $\langle \phi \rangle = 0$ in the minimum after inflation, and only $G_{\chi_{\pm}}^{\text{eff}}$ is nonzero. We will again assume that the third term in Eq.~\eqref{G_effective}) dominates and focus on the effects from a term $\delta K = a |H|^2 (Z^2 + \overline Z^2)/M_{Pl}^2$ in the K\"ahler potential. In total, we obtain
\bea 
| G_{\phi}^{\text{eff}} | &\sim& 0 \:, \\
|G_{\chi_{\pm}}^{\text{eff}}| &\sim& 3 a \frac{\langle\chi\rangle}{M_{P}} \frac{m_{3/2}}{m_{\chi_{\pm}}}\:. \label{geffective}
\eea
This means that only the decays of $\chi_{\pm}$ will produce gravitinos, while the gravitino production from inflaton decays will be practically zero.

\subsection{Estimation of the produced gravitino number density} \label{Estimation_of_the_produced_gravitino_number_density}

To estimate the gravitino number density we will make various simplifying approximations: We will begin our considerations when the fields  $\chi_{\pm}$ decay. We will assume that at this time the amplitude of the $\phi$ field oscillations is sufficiently small and that the energy density in the coherent oscillations simply redshifts like non-relativistic matter. In addition, we will divide the post-inflationary epoch in either radiation dominated or matter dominated phases. These approximations will allow us to derive rather simple analytical formulae. Of course, to calculate gravitino production in an explicit model numerically, one would drop these assumptions. 

We will furthermore assume that $\Gamma^{\chi}_{tot} > \Gamma^{\phi}_{tot}$, i.e.\ that $\chi_{\pm}$ decays earlier than the inflaton field $\phi$, and also that $m_{\chi_{\pm}} > 2 m_\phi$. This is generically true for most parameter choices in tribrid inflation consistent with observations (in particular for the parameter sets we use here as examples). 
 With $m_{\chi_{\pm}} > 2 m_\phi$, $\chi_{\pm}$ decays dominantly in two inflatons or its SUSY partners with total decay width approximately given by (see e.g.\ \cite{Antusch:2010mv})\footnote{In addition, $\chi_{\pm}$ decays (with lower rates) also to gravitinos as well as to MSSM particles.} 
\begin{equation} 
 \Gamma^{\chi}_{tot} \simeq \frac{\lambda^2 \kappa M^3}{8 \pi M_{P}^2}. \label{val}
\end{equation}

One important parameter for our analytic considerations will be the ratio of the energy densities stored in the $\chi_{\pm}$ field (in terms of homogenous scalar field oscillations) and $\phi$ field when $\chi_{\pm}$ decays, i.e.
\begin{equation}
\alpha = \tilde \rho_{\chi}/\tilde \rho_{\phi} \:,
\end{equation}
where the tilde denotes the time of $\chi_{\pm}$ decay. In the following we will treat $\alpha$ as a free parameter, as it depends on the details of the model and as its calculation would require the simulation of the preheating stage after tribrid inflation, which is beyond the scope of this paper.

At the time of $\chi_{\pm}$ decay all the decay products are relativistic and their energy densities therefore redshift as radiation. We can write for the energy densities $\tilde \rho_{\chi} =   \tilde \rho_{\chi}^{rad} + \tilde \rho_{3/2}$, where $ \tilde \rho_{3/2} \ll \tilde \rho_{\chi}^{rad}$, with $\tilde \rho_{\chi}^{rad}$ denoting all other radiation components from $\chi_{\pm}$ decay.

We will now estimate the gravitino production for different regimes of $\alpha$: 
\begin{itemize}
\item If $\alpha \leq 1$, the radiation energy density produced from $\chi_{\pm}$ decay is smaller than the energy density of the coherently oscillating $\phi$ field, which means that in our approximation we will treat the universe as matter dominated. It then remains matter dominated untill the $\phi$ field decays into radiation, and at this time we  calculate the reheat temperature. During the phase where the $\phi$ field dominates, the gravitinos from $\chi_{\pm}$ decay, which are treated as radiation, get diluted. 

\item On the other hand, if 
$\alpha \geq \sqrt{\Gamma^{\chi}_{tot}/\Gamma^{\phi}_{tot}}$ (as we will discuss below), the universe is (mainly) radiation dominated already after the time of $\chi_{\pm}$ decay (which is the case for $\alpha > 1$), and it will remain radiation dominated until the time of $\phi$ decay. The reheat temperature is then calculated at the time of $\chi_{\pm}$ decay. 

\item For intermediate values of $\alpha$, the universe is radiation dominated directly after $\chi_{\pm}$ decay, but becomes matter dominated before $\phi$ decay. In this case, the reheat temperature is calculated at the time of $\phi$ decay.
\end{itemize}

\noindent {\bf Case $\:\boldsymbol{\alpha \leq1}$:}
As mentioned above, in our approximation we will treat the universe as matter dominated in the epoch between $\chi_{\pm}$ decay and $\phi$ decay. The decay of the inflaton field $\phi$ dilutes the gravitino abundance produced from $\chi_{\pm}$ decay, and from $\phi$ decays no additional gravitinos are produced. The resulting gravitino abundance can be estimated as follows: 

The number density of gravitinos at the time of $\chi_{\pm}$ decay is given by 
\begin{equation}
\tilde n_{3/2} = \tilde n_{\chi}\frac{\Gamma^{\chi}_{3/2}}{\Gamma_{tot}^{\chi}}\cdot2,
\end{equation}
where the factor $2$ comes from the fact that $\chi_{\pm}$ decays into two gravitinos. For $\alpha \leq1$, the Hubble constant at the time of $\phi$ decay, $\hat H$, can be related to the Hubble constant at the time of $\chi_{\pm}$ decay, $\tilde H$, by
\begin{equation}
\frac{\tilde H}{\hat H} = \frac{\Gamma^{\chi}_{tot}}{\Gamma^{\phi}_{tot}} = \left(\frac{\tilde a}{\hat a}\right)^{-3/2}.
\end{equation}
So we obtain for the number density at the time of $\phi$ decay:
\begin{equation}
\hat n_{3/2} = \tilde n_{3/2} \left(\frac{\hat a}{\tilde a} \right)^{-3} = \tilde n_{3/2} \left(\frac{\Gamma^{\chi}_{tot}}{\Gamma^{\phi}_{tot}}\right)^{-2}.
\end{equation}
After $\phi$ decay, the gravitino number density normalised to the entropy density, $Y_{3/2}$  is thus given by
\begin{equation}
\hat Y_{3/2} = \frac{\hat n_{3/2}}{\hat s} = \frac{2 \alpha}{m_{\chi} } \left(\frac{\Gamma^{\chi}_{3/2}}{\Gamma^{\chi}_{tot}} \right)\frac{\hat \rho_{\phi}}{\hat s} \:,
\end{equation}
where $\tilde \rho_{\chi} = m_{\chi} \tilde n_{\chi}$ has been used. Assuming thermal equilibrium at the end of $\phi$ decay for simplicity, we use
\begin{equation}
\frac{\hat \rho_{tot}}{\hat s} = \frac{\hat \rho_{\phi}^{rad}}{\hat s}  + \frac{\hat \rho_{\chi}^{rad}}{\hat s}= \frac{\hat \rho_{rad}}{\hat s} = \frac{3}{4} T_{R},
\end{equation}
which leads to  
\begin{equation}
\frac{\hat \rho_{\phi}}{\hat s}  = \frac{3}{4} T_{R}\left(1 + \alpha\left(\Gamma^{\chi}_{tot}/\Gamma^{\phi}_{tot} \right)^{-2/3} \right)^{-1}.
\end{equation}
$T_{R}$ is the reheat temperature that has contributions from both $\phi$ and $\chi$ decays. Plugging in this relation gives us the following expression for the gravitino yield: 
\begin{equation} \label{yieldone}
\hat Y_{3/2} = \frac{2\alpha}{m_{\chi}} \left(\frac{\Gamma^{\chi}_{3/2}}{\Gamma^{\chi}_{tot}}\right) \frac{3 T_{R}}{4} \left(1 + \alpha \left(\Gamma^{\chi}_{tot}/\Gamma^{\phi}_{tot} \right)^{-2/3} \right)^{-1}.
\end{equation}
The second term in the bracket becomes negligible when either $\alpha \ll 1$, i.e.\ when most of the energy density after inflation is carried by the inflaton field $\phi$ (in which case there  is an overall suppression of $\hat Y_{3/2} $ by a factor $\alpha$), or when $\Gamma^{\chi}_{tot} \gg \Gamma^{\phi}_{tot} $, i.e.\ when $\phi$ decays much later than $\chi_{\pm}$. Under one of these conditions, the above formula reduces to the simple form
\begin{equation}
\hat Y_{3/2} = \frac{3}{2}\frac{\alpha}{m_{\chi_{\pm}}} \left(\Gamma^{\chi}_{3/2}/\Gamma^{\chi}_{tot} \right) T_{R}\:.
\label{Eq:gravitinio1}
\end{equation}

Interestingly, and in contrast to the case of ``conventional'' hybrid inflation, the gravitino yield is proportional to the reheat temperature $T_{R}$. Therefore by lowering $T_{R}$, which corresponds to the case that the inflaton field $\phi$ decays comparatively late, we can strongly suppress the non-thermally produced gravitino density $Y_{3/2}$ in tribrid inflation. We will now plug in values for the model parameters (as we did for ``conventional'' SUSY hybrid inflation) to check that for tribrid inflation the gravitino production can be sufficiently suppressed to avoid the ``gravitino problem''. 

For a rough estimate of the gravitino yield, we use $\Gamma^{\chi}_{tot} \simeq \Gamma^{\chi}_{\phi}$, neglecting the contribution from the decays to gravitinos and to Standard Model (MSSM) particles. The total decay rate of $\chi$ is then given by Eq.~\eqref{val}. Note that this is a conservative approximation, and will provide an estimate for an upper bound on $\hat Y_{3/2}$. Using Eq.~\eqref{geffective}, the gravitino yield is given by 
\begin{equation}
\hat Y_{3/2} = \frac{3}{8} \frac{\alpha a^2}{\kappa \lambda^2}\frac{m_{\chi}^2 \langle \chi \rangle^2}{M^3 M_{P}^2} T_{R}.
\end{equation}
Using the post-inflationary mass for the $\chi_\pm$ fields, $m_{\chi_{\pm}}^2 \simeq 4 \kappa^2 M^2$, as in Eq.~\eqref{yield_hybrid} for the case of ``conventional'' SUSY hybrid inflation, we find the corresponding expression for tribrid inflation 
\begin{equation} \label{yield_tribrid}
m_{LSP} Y_{3/2}^{\phi} \simeq \frac{4\alpha a^2 \kappa}{\lambda^2} \!\left(\frac{\langle \chi \rangle}{10^{16}\,\text{GeV}} \right)^2 \!\!\left(\frac{5\cdot10^{15}\,\text{GeV}}{M} \right) \!\left(\frac{m_{LSP}}{100\,\text{GeV}}\right) \!\left(\frac{T_R }{10^{7}~\text{GeV}} \right)\!10^{-12}\,\text{GeV},
\end{equation}
Plugging in the typical values for tribrid inflation $ \langle \chi \rangle \sim 10^{16}$ GeV, $M = 5 \cdot 10^{15}$ GeV, $\kappa = 0.1, \lambda = 0.1$ \cite{Antusch:2008pn,Antusch:2009ef}, we see that the abundance is well bellow the observational limit of Eq.~\eqref{gravitino_constraint} for $m_{LSP} \sim 100$ GeV and for the example reheat temperature $T_{R} \lesssim 10^{7}$ GeV. 

Note that for reheat temperatures around $10^{7}\:\mbox{GeV}$ the thermal production of gravitinos can indeed be neglected. We like to emphasize that, in contrast to the non-thermal gravitino production in standard SUSY hybrid inflation, the gravitino abundance in tribrid inflation is proportional to the reheat temperature $T_{R}$. 
Thus by lowering $T_{R}$, in tribrid inflation models one is also reducing the non-thermal gravitino production.

\vspace{0.5cm}
\noindent {\bf Case $\:\boldsymbol{ 1 < \alpha < \sqrt{\Gamma^{\chi}_{tot}/\Gamma^{\phi}_{tot}}}\,$:}

\begin{figure}
  \centering
   \includegraphics[width=0.6\textwidth]{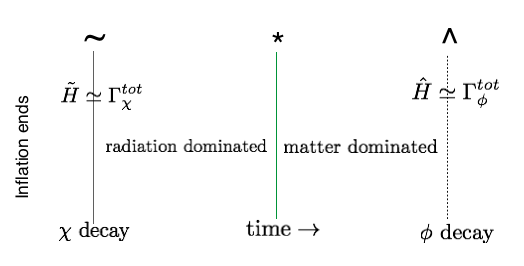} 
 \caption{The diagram illustrates the different times and epochs used in the calculation of our estimates for the gravitino number density $\hat Y_{3/2}$ in tribrid inflation. For $\alpha \rightarrow 1$, the time $t^*$, where radiation and matter energy densities are equal (shown by the green line in the diagram) is moved further towards the left and merges with $\tilde t$, the time when $\chi_{\pm}$ decays. For $\alpha \rightarrow \sqrt{\Gamma^{\chi}_{tot}/\Gamma^{\phi}_{tot}}$, $t^*$ moves towards the right and merges with the time $\hat t$ when $\phi$ decays. 
 }
  \label{fig:cartoon}
\end{figure}

If $1 < \alpha < \sqrt{\Gamma^{\chi}_{tot}/\Gamma^{\phi}_{tot}}$, the decay of the $\chi$ field would produce a radiation dominated era which would be followed by a matter dominated phase with coherent oscillations of the $\phi$ field. The time at which radiation and matter densities become equal, will be denoted by ``$*$'' (cf.\ Fig.~\ref{fig:cartoon} for a schematic illustration). Using $H \propto a^{-2}$ for the radiation dominated epoch and $H \propto a^{-3/2}$ for the time of matter domination, we arrive at
\begin{equation}
a_{*}/\hat a = \alpha^{4/3} \left(\Gamma^{\chi}_{tot}/\Gamma^{\phi}_{tot}  \right)^{-2/3} \:, 
\end{equation}
where $H_{*} = \Gamma^{\chi}_{tot}(\tilde a/a_*)^2$ has been used. This translates to the following condition for having both, a radiation dominated and a matter dominated phase between $\chi_{\pm}$ and $\phi$ decay: 
\begin{equation}
1 < \alpha < \sqrt{\Gamma^{\chi}_{tot}/\Gamma^{\phi}_{tot}}\:.
\end{equation}
Again, tracking the energy densities during different epochs we arrive at
\begin{equation}
\hat a/\tilde a = \alpha^{-1/3} \left(\Gamma^{\chi}_{tot}/\Gamma^{\phi}_{tot} \right)^{2/3}\:,
\end{equation}
where $\alpha = \tilde \rho_{\chi}/\tilde \rho_{\phi} = a_*/\tilde a$ has been used. 
Following similar steps as for the case $\alpha \leq1$, we find that the gravitino number density $\hat Y_{3/2}$ can be estimated as
\begin{equation} \label{yield_inter}
Y_{3/2} = \frac{2\alpha}{m_{\chi}} \left(\frac{\Gamma^{\chi}_{3/2}}{\Gamma^{\chi}_{tot}}\right) \frac{3 T_{R}}{4} \left(1 + \alpha^{4/3}\left(\Gamma^{\chi}_{tot}/\Gamma^{\phi}_{tot} \right)^{-2/3} \right)^{-1}.
\end{equation}
We see that (as it should be) for $\alpha = 1$ the above formula matches to the one in \eqref{yieldone}.

\vspace{0.5cm}
\noindent {\bf Case $\:\boldsymbol{\alpha \geq \sqrt{\Gamma^{\chi}_{tot}/\Gamma^{\phi}_{tot}}}$:}

In the limit $\alpha \rightarrow \sqrt{\Gamma^{\chi}_{tot}/\Gamma^{\phi}_{tot}}$, the time $t^*$ where radiation and matter energy densities are equal merges with the time $\hat t$ when $\phi$ decays. This means that there will be no phase of intermediate matter domination and thus in our approximation we treat the universe as radiation dominated from the time of $\chi_{\pm}$ decay on. Consequently, in our approximation, there will be no further dilution of the gravitinos produced from $\chi_{\pm}$ decay.
Furthermore, the reheat temperature is now defined at the time of $\chi_{\pm}$ decay. We therefore expect that in some approximation one can calculate the gravitino abundance as in ``conventional'' SUSY hybrid inflation from $\chi_{\pm}$ decay, i.e.\ using the formula from Eq.~\eqref{yield_basic} with $\phi$ replaced by $\chi$:
\begin{equation}\label{eq:matched}
Y_{3/2} \simeq 2\frac{\Gamma^\chi_{3/2}}{\Gamma^{\chi}_{tot}} \frac{3}{4}\frac{T^\chi_{R}}{m_{\chi_{\pm}}}.
\end{equation}

To see that is indeed true, let us consider the limiting case $\alpha = \sqrt{\Gamma^{\chi}_{tot}/\Gamma^{\phi}_{tot}}$  in more detail: For this value, $\hat\rho^{rad}_\phi = \hat\rho^{rad}_\chi$, and the reheat temperature, $T_{R} = \frac{4}{3}(\hat\rho^{rad}_\phi + \hat\rho^{rad}_\chi)/\hat s$ can be written as $T_{R} = \frac{4}{3}(2 \hat\rho^{rad}_\phi)/ \hat s =: 2 \, T^\phi_{R}$,  defining $T^\phi_{R} := \frac{4}{3}\hat\rho^{rad}_\phi/\hat s$ as the reheat temperature calculated from $\phi$ decay alone (ignoring $\hat\rho^{rad}_\chi$). Defining analogously $T^\chi_{R} := \frac{4}{3}\tilde\rho^{rad}_\chi/\hat s$ (at the time of $\chi_\pm$ decay), using 
\begin{equation}
T^\chi_{R}/T^\phi_{R} = \sqrt{\Gamma^{\chi}_{tot}/\Gamma^{\phi}_{tot}} \:,
\end{equation}
and plugging in $\sqrt{\Gamma^{\chi}_{tot}/\Gamma^{\phi}_{tot}} = \alpha$, we indeed verify that Eq.~\eqref{yield_inter} reduces to Eq.~(\ref{eq:matched}). Note that using $T^\chi_{R}$ in Eq.~(\ref{eq:matched}) means that the contribution $\tilde\rho_\phi$ is neglected when calculating the reheat temperature at the time of $\chi_\pm$ decay (which is the approximation in which both formulae match).  

Let us note at this point that $\Gamma^{\chi}_{tot}/\Gamma^{\phi}_{tot}$ is actually a quite large number. Considering again the limiting case $\sqrt{\Gamma^{\chi}_{tot}/\Gamma^{\phi}_{tot}} = \alpha$ where $T^\phi_{R} = \tfrac{1}{2} T_{R}$, and plugging in the above used example values for the tribrid inflation model parameters $ \langle \chi \rangle \sim 10^{16}$ GeV, $M = 5 \cdot 10^{15}$ GeV, $\kappa = 0.1, \lambda = 0.1$ \cite{Antusch:2008pn,Antusch:2009ef}, we obtain (with $\Gamma^{\chi}_{tot}$ from Eq.~\eqref{val} and $\Gamma^{\phi}_{tot}$ from Eq.~\eqref{eq:TR_Phi}): 
\begin{equation}
\frac{\Gamma^{\chi}_{tot}}{\Gamma^{\phi}_{tot}} = \frac{2\,M^3}{T_R^2 \,M_P} \frac{\lambda^2 \kappa}{8 \pi}\left(\frac{\pi^2 g_*}{10}\right)^{-\tfrac{1}{2}} \!\simeq \:3\cdot 10^9  \left(\frac{10^{7}\:\text{GeV}}{T_R } \right)^2  \:.
\end{equation}
This implies in particular that $\alpha$ is very large, $\alpha = \sqrt{\Gamma^{\chi}_{tot}/\Gamma^{\phi}_{tot}} \simeq 5 \cdot 10^4$ for $T_R \simeq 10^7$ GeV, which means that it is a very good approximation to neglect $\tilde\rho_\phi$ when calculating the reheat temperature at the time of $\chi_\pm$ decay (since $\tilde\rho_\chi = \alpha \tilde\rho_\phi \gg \tilde\rho_\phi $). 
On the other hand, we note that we do not expect that  
such a large $\alpha$, where $\alpha \geq \sqrt{\Gamma^{\chi}_{tot}/\Gamma^{\phi}_{tot}}$, can be realised. Therefore, the non-thermally produced gravitino abundance in tribrid inflation should be generically suppressed.

\section{Conclusions} \label{conclusion}
In this paper, we have investigated non-thermal gravitino production after tribrid inflation. Tribrid inflation is a variant of hybrid inflation where the inflaton resides in the matter sector of the theory (cf.\ Sec.~\ref{Tribrid_inflation_vs_conventional_SUSY_hybrid_inflation}). We find that in contrast to conventional supersymmetric hybrid inflation, where non-thermal gravitino production imposes severe constraints on the inflationary model, the ``non-thermal gravitino problem'' is generically absent in models of tribrid inflation. 

As in ``conventional'' SUSY hybrid inflation, at the end of tribrid inflation the dynamics is governed by oscillations and decays of the inflaton field and the waterfall field. However, in contrast to ``conventional'' SUSY hybrid inflation, these fields are not degenerate in mass but (in the considered tribrid inflation setup) the inflaton field is much lighter than the waterfall field. While the decays of the waterfall field in tribrid inflation (and of both fields in ``conventional'' SUSY hybrid inflation) produce gravitinos as discussed in Sections~\ref{Non-thermal_gravitino_production_in_hybrid_inflation} and \ref{grav_production_tribrid}, the inflaton field in tribrid inflation has a negligible decay rate into gravitinos. The two main effects which allow to easily solve the ``non-thermal gravitino problem'' in tribrid inflation are: 

\begin{itemize}
\item With the inflaton being lighter than the waterfall field, in tribrid inflation the latter automatically has a second decay channel with a much larger rate than for the decay into gravitinos. This leads to a strongly suppressed gravitino production from the decays of the waterfall field. 
\item The inflaton in the considered tribrid inflation setup generically decays later than the waterfall field. While it does not produce gravitinos from its decays (cf.\ subsection~\ref{Decay_rates_into_gravitinos}), it even dilutes the gravitino population produced earlier from the waterfall field decays. The later the inflaton decays, the smaller the reheat temperature $T_R$, and the stronger is the dilution effect. 
\end{itemize}

To quantify how much the ``non-thermal gravitino problem'' is relaxed, in subsection~\ref{Estimation_of_the_produced_gravitino_number_density} we calculated estimates for different regimes of $\alpha$, which parameterizes the ratio of the energy densities in the waterfall field and in the inflaton field after inflation. Comparing the produced gravitino abundances with conventional F-term SUSY hybrid inflation, we find that the overall amount of produced gravitinos is in general much lower (cf.\ e.g.\ Eqs.~\eqref{yield_hybrid} vs.\ \eqref{yield_tribrid}), and also that there is a different dependence on the model parameters. For instance, in tribrid inflation a lower $T_R$ allows to suppress the produced gravitino abundance, whereas in conventional F-term SUSY hybrid inflation, the gravitino yield is proportional to $1/T_R$.

In summary, models of tribrid inflation not only offer the attractive possibility to identify the inflaton with a field (or a $D$-flat direction of fields) from the matter sector of the theory. As we showed in this letter, they also allow to solve the ``non-thermal gravitino problem'' easily due to some built-in model properties.  

\section*{{Acknowledgements}}

We thank Sebastian Halter for collaboration at the early stage of the project.
KD would like to thank  the Max Planck Institute for Physics, Munich, where the work was initiated and in part completed. KD is partially supported by a Ramanujan fellowship and a DST-Max Planck Society visiting fellowship.

\end{document}